\documentstyle[twoside,fleqn,espcrc2,epsf]{article}

\newcommand{\Z}{{Z \!\!\! Z}}
\newcommand{\dual}{\mbox{}^{\ast}}
\newcommand{\LL}{{I\!\! L}}

\newcommand{\eq}[1]{(\ref{#1})}

\newcommand{\beq}{\begin{equation}}
\newcommand{\eeq}{\end{equation}}
\newcommand{\beqn}{\begin{eqnarray}}
\newcommand{\eeqn}{\end{eqnarray}}

\newcommand{\cD}{{\cal D}}

\def\cC{{\cal C}}

\newcommand{\cZ}{{\cal Z}}

\newcommand{\intpiD}{\int\limits_{-\pi}^{+\pi} {\cD}}

\newcommand{\intinfD}{\int\limits_{-\infty}^{+\infty} {\cD}}

\newcommand{\CK}[1]{\mbox{\scriptsize c}_{\mbox{$\scriptstyle #1$}}}
\newcommand{\nsum}[2]{\sum_{ #1(\CK{#2}) \in \Z }}

\newcommand{\nddsum}[2]{\sum_{\stackrel{\scriptstyle \dual #1(\dual\CK{#2})
\in \Z} {\delta \dual #1=0}}}
\newcommand{\dd}{\mbox{d}}

\newcommand{\AmS}{{\protect\the\textfont2
 A\kern-.1667em\lower.5ex\hbox{M}\kern-.125emS}}

\def\dd{{\rm d}}

\def\NP{ Nucl.~Phys.}

\def\PRL{ Phys.~Rev.~Lett.}

\title{\vspace{-3.6cm}
\begin{flushright}
{\normalsize
ITEP-TH-37/96\\
\vspace{-.2cm}
August 1996}
\end{flushright}
\vspace{1.5cm}
Aharonov--Bohm Effect in 3D Abelian Higgs 
Theory\thanks{Talk given at Lattice 96, St.~Louis, USA.}}

\author{M.N.~Chernodub, F.V.~Gubarev and M.I.~Polikarpov
       \address{ITEP, B.Cheremushkinskaya 25, Moscow,
       117259, Russia}}

\begin{document}

\begin{abstract}
We study a field--theoretical analogue of the Aharonov--Bohm effect
in the 3D Abelian Higgs Model: the corresponding topological
interaction is proportional to the linking number of the vortex and
the particle world trajectories.  We show that the Aharonov--Bohm
effect gives rise to a nontrivial interaction of tested charged
particles.
\end{abstract}

\maketitle

\section{Introduction}

It is well known that the Abelian Higgs Model in three and four
dimensions has the classical solutions called
Abrikosov--Nielsen--Olesen vortices~\cite{AbNiOl}. These vortices
carry quantized magnetic flux (vorticity) and the wave function of
the charged particle which is scattered on the vortex acquires an
additional phase. The shift in the phase is the physical effect which
is the field--theoretical analog \cite{ABE} of the
quantum--mechanical Aharonov--Bohm effect \cite{AhBo59}: vortices
play the role of solenoids, which scatter the charged particles.

In Section~2 we show that in the $3D$ lattice Abelian Higgs model with
the non--compact gauge field, the Aharonov--Bohm effect gives rise to the
long--range Coulomb--like interaction between test particles. In the
three dimensional case the induced potential is confining, since the
Coulomb interaction grows logarithmically.

In Section~3 we present the results of the numerical calculations of
the induced potential. Our numerical results show the existence of
the Aharonov--Bohm effect in this model.

\section{Potential Induced by the Aharonov--Bohm Effect}

The partition function of $3D$ non--compact lattice Abelian Higgs
Model~is:

\beqn
  \cZ = \intinfD A \intpiD \varphi \nsum{l}{1} e^{ - S(A,\varphi,l)}
  \,,\label{NC}
\eeqn
where
\beqn
  S(A,\varphi,l) = \beta \|\dd A\|^2 +
  \gamma \| \dd\varphi + 2\pi l - N A \|^2\,,
  \label{ActionNC}
\eeqn
$A$ is the non--compact gauge field, $\varphi$ is the phase of the
Higgs field and $l$ is the integer--valued one--form. For simplicity
we consider the limit of the infinite Higgs boson mass, then the
radial part of the Higgs field is frozen and we use the Villain form
of the action.

One can rewrite the integral \eq{NC} as the sum over the closed
vortex trajectories using the analogue of
Berezinski--Kosterlitz--Thauless (BKT) transformation
\cite{BKT}\footnote{For D=4 the similar transformation was performed
in ref.\cite{PoWiZu93}, we use the notations of this paper}:

\beq
  \cZ \propto \cZ^{BKT} =  \hspace{-2mm} \nddsum{j}{1}
  \hspace{-2mm} e^{ - 4 \pi^2 \gamma
  \left(\dual j, {(\Delta + m^2)}^{-1} \dual j \right)}
  \,, \label{TD}
\eeq
where $m^2 = N^2 \gamma \slash \beta$ is the $classical$ mass of the
vector boson $A$. The closed currents $\dual j$ which are defined on
the dual lattice represent the vortex trajectory.
It can be easily seen from eq.\eq{TD}
that the currents $\dual j$ interact with each other through the
Yukawa forces.

In the limit $N^2 \gamma \gg 1 \gg \beta$ the partition function
\eq{TD} becomes:

\beqn
  \cZ^{BKT}_0 = \nddsum{j}{1}
  \exp\biggl\{- \frac{4 \pi^2 \beta}{N^2}
  {||\dual j||}^2 \biggr\}\,.
 \label{TD2}
\eeqn
In the corresponding continuum theory the term
${||\dual j||}^2$ is proportional to the length of the trajectory
$\dual j$, therefore the vortices are free. The quantum average of
the Wilson loop $W_M(\cC) = \exp\{ i M (A,j_\cC) \}$ in the discussed
region of the parameters is:

\beqn
 {<W_M(\cC)>}_N = \frac{1}{\cZ^{BKT}_0}
 \nddsum{j}{1} \nonumber\\
 \exp\biggl\{- \frac{4 \pi^2 \beta}{N^2} {||\dual j||}^2 +
 2 \pi i \frac{M}{N} \LL\left(\dual j,j_\cC\right) \biggr\}\,.
 \label{wl2}
\eeqn
The last long--range term has the topological origin: $\LL (\dual
j,j_\cC)$ is the linking number between the world trajectories
of the defects $\dual j$ and the Wilson loop~$j_\cC$:

\beq
       \LL(\dual j,j_\cC) = (\dual j_\cC, {\Delta}^{-1} \dd \dual
       j)\,.  \label{Ll}
\eeq
The trajectory of the vortex $\dual j$ is a closed loop and the
linking number $\LL$ is equal to the number of points at which the
loop $j_\cC$ intersects the two dimensional surface bounded by the
loop $\dual j$. The equation \eq{Ll} is the lattice analogue of the
Gauss formula for the linking number. This topological interaction
corresponds to the Aharonov--Bohm effect in the field
theory~\cite{ABE,Trento}.  Thus, eq.\eq{wl2}
describes the Aharonov--Bohm interaction of the free vortices
carrying the flux $\frac{2 \pi}{N}$ with the test particle
of the charge $M$. Therefore the interaction between the charged
particles is due to Aharonov--Bohm effect $only$.

The estimation of \eq{wl2} in the saddle--point approximation
($N^2 \slash \beta \gg 1$) gives in the leading order~\cite{Trento}:

\beq
  {<W_M(\cC)>}_N = const. e^{ -
  \kappa^{(0)}_{(M, N)} \cdot
  \left(j_\cC, \Delta^{-1} j_\cC\right)}\,,
  \label{wl3}
\eeq
where

\beqn
  \kappa^{(0)}_{(M, N)} = \frac{q^2 N^2}{4 \beta}\,, \qquad
  q = \min_{K \in \Z} |\frac{M}{N} - K|\,,
  \label{q}
\eeqn
$q$ is the distance between the ratio $M \slash N$ and a nearest
integer number. Expressions (\ref{wl3}--\ref{q}) depend on the
fractional part of $M \slash N$, this is the consequence of the
Aharonov--Bohm effect. Interaction of the testing charges is absent
if $q = 0$ ($M \slash N$ is integer), this corresponds to the
complete screening of the test charge $M$ by the Higgs
bosons of the charge $N$.

Consider the product of two Polyakov lines: $W_M(\cC) =
L^+_M(0) \cdot L_M(R)$. Then $(j_\cC, \Delta^{-1} j_\cC) = 2 T \,
\Delta^{-1}_{(2)}(R)$, where $\Delta^{-1}_{(2)}(R)$ is the
two--dimensional massless lattice propagator. At large $R$ the
propagator $\Delta^{-1}_{(2)}(R)$ grows logarithmically,
$\Delta^{-1}_{(2)}(R) = \frac{C_0}{2} \ln R + \dots$, where $C_0$ is
some numerical constant. Then eq.\eq{wl3} is reduced to:

\beqn
  {<L^+_M(0) L_M(R)>}_N = const. \, e^{ - V_{(M,N)}(R)}\,,
  \label{VMN}
\eeqn
where

\beqn
 V_{(M,N)}(R) = C_0 \, \kappa^{(0)}_{(M, N)} \cdot \ln R + \cdots\
 \label{wlR}
\eeqn
is the long--range potential induced by the Aharonov--Bohm effect.

One can prove the following general statement: if the potential
$V_{(M,N)}$ is induced by the Aharonov--Bohm effect, then it must
satisfy the following relations:

\beqn
  V_{(M,N)} = V_{(N - M,N)}\,, \qquad V_{(N,N)} = 0\, ,
  \label{gen}
\eeqn
see {\it e.g.} the definition of $q$ \eq{q}.

\section{Numerical Calculations}

We calculated numerically the potential between the tested particles
with the charge $M$ in the three dimensional Abelian Higgs Model, the
charge of the Higgs boson is $N=6$. The action of the model is chosen
in the Wilson form:  $S[A,\varphi] = \beta {||\dd A||}^2 - \gamma
\cos (\dd \varphi + N A)$. In our calculations the standard
Monte--Carlo method is used. The simulations are performed on the
lattice of the size $16^3$ for the charges $M = 1, \dots, N$.

We fit the numerical data for the potential,
\eq{VMN}, by the formula:

\beq
  V_{(M,N)}(R) = 2 \, \kappa^{num}_{M} \cdot
  T \cdot \Delta^{-1}_{(2)}(R) + C^{num}\,,
  \label{fit}
\eeq
where $\kappa^{num}$ and $C^{num}$ are numerical fitting parameters.
It turns out that the numerical data for $\kappa^{num}_{M}$ are
well described by the formula~\eq{fit}.

\begin{figure}[bth]
~
\vspace{-1.3cm}
\centerline{\epsfxsize=0.40\textwidth\epsfbox{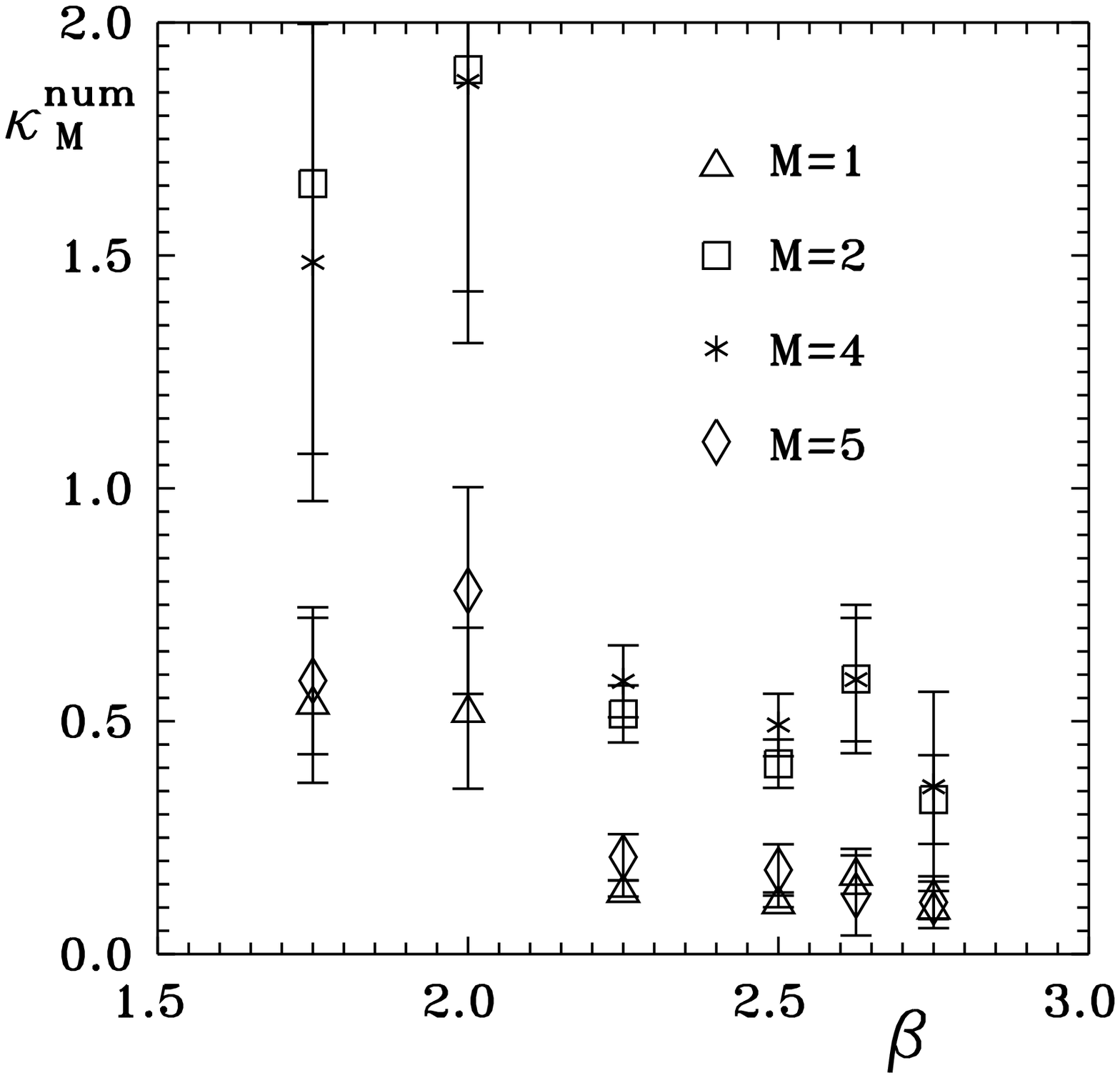}}
\vspace{0.7cm}
\centerline{Figure 1. The coefficients $\kappa^{num}_{M}$
$vs.$ $\beta$.}
~
\vspace{-1.2cm}
\end{figure}

We present on the Figure~1 the plot of the dependence of the
coefficients $\kappa^{num}_{M}$ on $\beta$ for $M=1,2,4,5$ and
$\gamma = 15$.  It's seen, that $\kappa^{num}_{M} = \kappa^{num}_{N -
M}$ within the numerical errors. The potential for the charge $M=N$
is equal to zero within the errors. These relations are in agreement
with the Aharonov--Bohm nature of the potential $V_{(M,N)}(R)$, {\it
cf.} eq.\eq{gen}.

The Abelian Higgs model is in the Coulomb phase at the considered
values of the parameters $\beta$ and $\gamma$.
The usual Coulomb interaction of the test particles with the charge
$M$ is proportional to $M^2$, the fact that the potential $V_{(M)}$
satisfies the relations \eq{gen} also means that the Coulomb
interaction of the tested particles is small.

Although both theoretical and measured numerical potentials satisfy
eq.\eq{gen} the measured coefficients $\kappa^{num}_M(\beta)$ are not
described by semiclassical formula \eq{q}. This deviation is
due to the renormalization of $\kappa^{(0)}$ by quantum
corrections. The same effect exists in the considered model at the finite
temperature~\cite{GuPoCh96}.

\section*{Conclusion and Acknowledgments}

In this talk we show both analytically and numerically that the
Aharonov--Bohm interaction between the vortices and the charged
particles induce the long--range Coulomb--type interaction between
the particles. Due to the long--ranged nature of the induced
potential the Aharonov--Bohm effect may play a role in the dynamics
of colour confinement in nonabelian gauge theories~\cite{ChPoZu94}.

This work is supported by the JSPS Program on Japan -- FSU scientists
collaboration, by the Grants INTAS-94-0840, INTAS-94-2851, and by
Grant No. 96-02-17230a, financed by the Russian Foundation for
Fundamental Sciences.

\end{document}